# Magneto-optical study of ZnO based diluted magnetic semiconductors

K.Ando*

Nanoelectronics Research Institute, National Institute of Advanced Industrial Science and Technology

[Abstract]

Magneto-optical properties of ZnO:Co and ZnO:Ni films were measured. Magnetization measurements show that some of the films are paramagnetic and others are ferromagnetic. Magnetic circular dichroism clarified that $Zn_{1-x}Co_xO$ and $Zn_{1-x}Ni_xO$ included in samples are paramagnetic diluted magnetic semiconductors. Ferromagnetic precipitations seem to be responsible for the observed ferromagnetic behaviors. Criteria to judge the ferromagnetic DMS are also discussed.

## 1. Introduction

Ferromagnetic orderings in ZnO-based diluted magnetic semiconductors (DMSs) were theoretically predicted by Sato *et al.*[1] and Dietl *et al.* [2]. Ando *et al.* [3,4] investigated the magneto-optical properties of $Zn_{1-x}TM_xO$ (TM = Sc, Ti, V, Cr, Mn, Co, Ni, Cu) films[5,6], and found that the delocalized *sp*-electrons interact with the localized *d*-electrons in this system. But they have not observed any ferromagnetic ordering down to 5K. On the contrary, Ueda *et al.* [7] reported ferromagnetic behaviors of Co-doped ZnO films. The Curie temperature is higher than room temperature. Wakano *et al.* [8] also reported the ferromagnetic behaviors of Ni-doped ZnO films.

In this report, we discuss the ferromagnetic ordering in the ZnO:Co and ZnO:Ni from their magneto-optical properties.

## 2. Experiment

All films used in this study were grown on sapphire (0001) substrates. Magnetic properties were estimated by a superconducting quantum interface device (SQUID). $Zn_{1-x}Co_xO$ and $Zn_{1-x}Ni_xO$ films grown at Tokyo Institute of Technology were paramagnetic [9]. A ZnO:Co(15%) film grown at Osaka University showed a clear magnetic hysteresis loop. The magnetic moment at T =10K was about 0.2 $\mu_B$ per Co ion, which is one order smaller than that of the ferromagnetic ZnO:Co film used for Ref.7. A ZnO:Ni(1%) film grown at Osaka Prefecture University also showed ferromagnetic behaviors at 5K and 300K [8].

Magnetic circular dichroism (MCD), which detects the relative difference of the circular polarization dependent optical absorption, were measured with applied magnetic fields along the light propagation direction [10].

### 3. Co-doped ZnO

Figure 1 shows a MCD spectrum of the paramagnetic $Zn_{0.9}Co_{0.1}O$ film. The MCD intensity changes linearly with magnetic field. MCD spectral shape near 2.0eV is a fingerprint of *d-d\** intra-ionic transitions from $^4A_2$ (*F*) to $^4T_1(P)$, $2E(G)$ of $Co^{2+}$ ions situated in the $T_d$ symmetry sites. This indicates a substitutional replacement of the tetrahedrally coordinated $Zn^{2+}$ ions by $Co^{2+}$ ions in $Zn_{1-x}Co_xO$. The structures observed near 3.4eV are due to the optical transition from the valence band to the conduction band of $Zn_{1-x}Co_xO$ [3].

Figure 2 shows a MCD spectrum of the ferromagnetic ZnO:Co film. In addition to the sharp structures near 2eV and 3.4eV, a broad MCD signal appeared. The magnetic field dependence of the MCD intensity at 1.90eV is shown in Fig.3. A ferromagnetic hysteresis appeared, and the MCD changed its polarity with magnetic field. This strange magnetic field dependence means that the observed MCD signal is composed of ferromagnetic and paramagnetic components. Since the filed dependence of the MCD is linear in the high field region, we can easily decompose the MCD spectrum into the paramagnetic and ferromagnetic components from the MCD spectra measured at 5kOe and 10kOe. The spectral shape of the paramagnetic MCD component (Fig.4) of the ferromagnetic ZnO:Co sample is same as that of the paramagnetic $Zn_{0.9}Co_{0.1}O$ (Fig.1). This means that the paramagnetic $Zn_{1-x}Co_xO$ DMS was contained in the ferromagnetic ZnO:Co sample. The spectral shape of the ferromagnetic MCD component (Fig.5) of the ferromagnetic ZnO:Co sample did not show noticeable structures near 2eV and 3.4eV, which are the fingerprint energies of the band structures of the host ZnO semiconductor. The electronic structure of the ferromagnetic material contained in the ferromagnetic ZnO:Co sample seems to have no relation with that of ZnO. The broadened MCD spectral shape suggests that this material has a metallic nature. This ZnO:Co sample shows the same broadened MCD spectrum at 300K.

These results suggest that some ferromagnetic precipitations are easily formed in ZnO:Co sample. At present, it is not clear if the strong ferromagnetism of ZnO:Co reported in Ref.7 is also due to the precipitations or not.

### 4. Ni-doped ZnO

Figure 6 shows a MCD spectrum of the paramagnetic $Zn_{0.97}Ni_{0.03}O$ film. The MCD intensity changes linearly with magnetic field.

Figure 7 shows a MCD spectrum of the ferromagnetic ZnO:Ni film. Its spectral shape is identical with that of the paramagnetic $Zn_{0.97}Ni_{0.03}O$ (Fig.6). The MCD intensity changes linearly with magnetic field (Fig.8). The ferromagnetic behaviors of the ZnO:Ni sample comes from some ferromagnetic materials which do not have magneto-optical activity. Because the strong magneto-optical activity due to the *sp-d* exchange interactions is the most distinguishing characters of DMS, the ferromagnetic materials included in the ferromagnetic ZnO:Ni sample seem not to be DMS.

**5. How to confirm the ferromagnetism of diluted magnetic semiconductors**

As discussed above, some ferromagnetic precipitations are easily formed in ZnO:TM films. In general, the crystallographical tools such as the X-ray diffraction, RHEED diffraction, and TEM observations are too poor to detect the precipitations. On the contrary, SQUID magnetization measurements are so sensitive to be disturbed by the ferromagnetic precipitations. Therefore ***we should not rely on the crystallographical results and magnetic data for judging the ferromagnetism of DMSs.*** It should be reminded that the most distinguishing character of DMSs is the strong *sp-d* exchange interaction. No material can be called as DMS unless the delocalized *sp*-carriers is confirmed to be strongly influenced by the external magnetic field through the *d*-electrons. When the way of the influence is of the ferromagnetic type, the material is a ferromagnetic DMS.

The best tool for such examinations is the magneto-optical characterization such as the Zeeman-split observation and the MCD measurement [10]. The Hall effect measurement is also useful. But the magneto-optical measurements are more reliable because their spectral shapes also give us the fingerprints of the material.

The MCD measurements have confirmed the intrinsic ferromagnetism in $Ga_{1-x}Mn_xAs$[11], $In_{1-x}Mn_xAs$[12], and $Zn_{1-x}Cr_xTe$[13]. Recently, several new materials, such as $TiO_2$:Co[14], GaN:Mn[15], and (Cd,Ge)$P_2$:Mn[16], were reported to be ferromagnetic DMSs. These reports are based on the crystallograhic and magnetic data. It is desired to confirm their ferromagnetism by the magneto-optical characterization.

Acknowledgements

The author thanks Prof.M.Kawasaki and Dr.T.Fukumura of Tokyo Institute of Technology, Prof.H.Tabata of Osaka University, and Prof.N.Fujimura of Osaka Prefecture University for supplying their samples.

*ando-koji@aist.go.jp

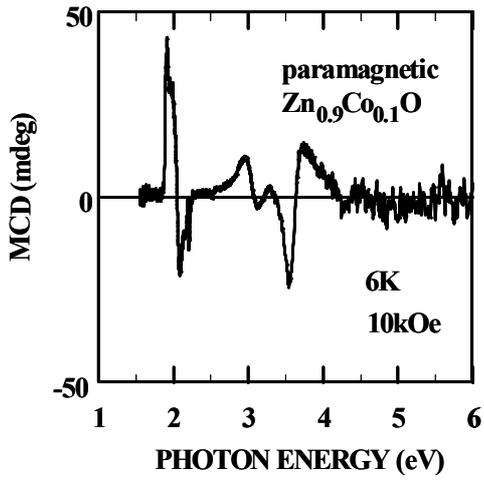

[Fig.1]

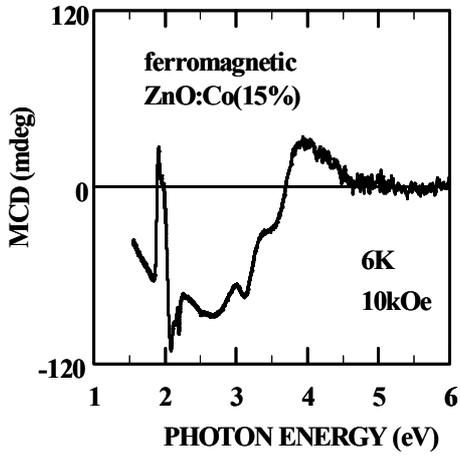

[Fig. 2]

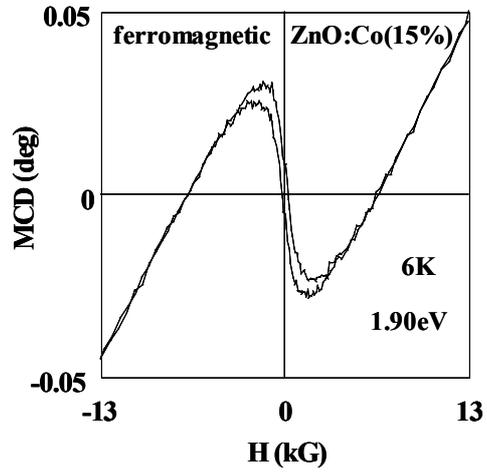

[Fig. 3]

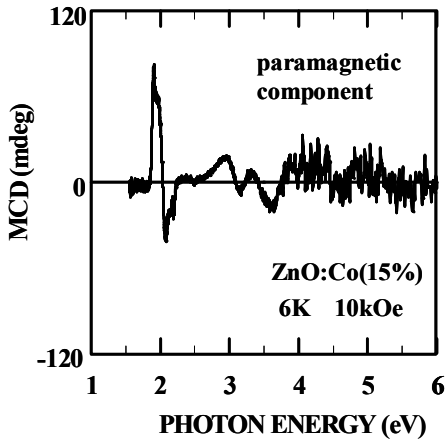

[Fig. 4]

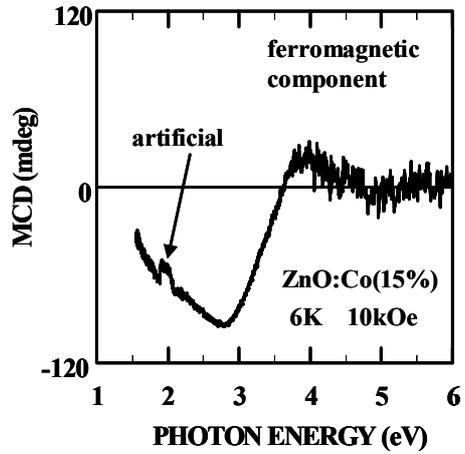

[Fig. 5]

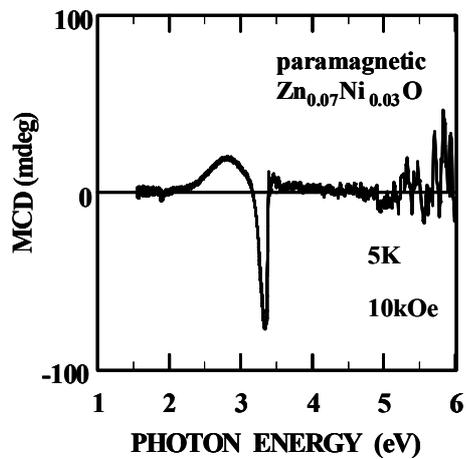

[Fig. 6]

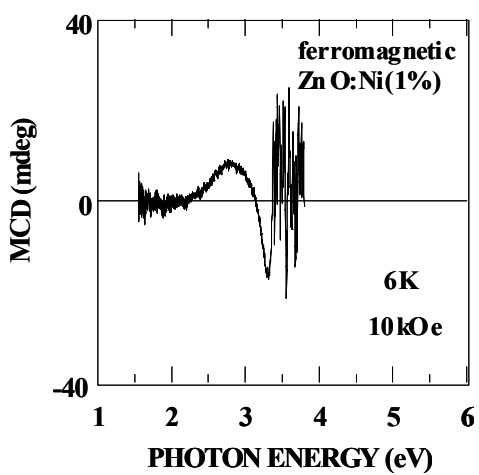

[Fig. 7]

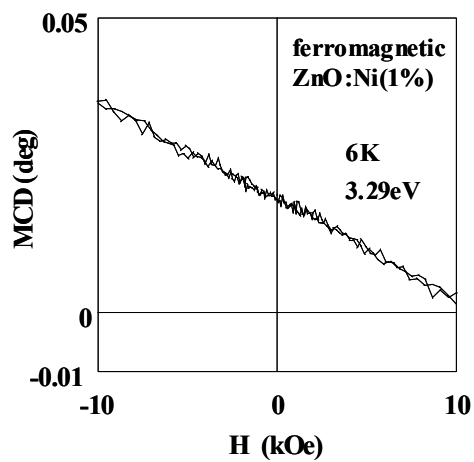

[Fig. 8]